\documentclass{aa}  
\usepackage{graphicx}
\usepackage{txfonts}
\usepackage[breaklinks=true]{hyperref}
\usepackage{ctable}
\usepackage{mathtools}

\usepackage{soul}

\usepackage{natbib,twoopt}
\usepackage[breaklinks=true]{hyperref} 
\hypersetup{colorlinks=true, linkcolor=blue, urlcolor=blue, citecolor=blue}
\bibpunct{(}{)}{;}{a}{}{,} 
\makeatletter
\newcommandtwoopt{\citeads}[3][][]{\href{http://adsabs.harvard.edu/abs/#3}%
{\def\hyper@linkstart##1##2{}%
\let\hyper@linkend\@empty\citealp[#1][#2]{#3}}}
\newcommandtwoopt{\citepads}[3][][]{\href{http://adsabs.harvard.edu/abs/#3}%
{\def\hyper@linkstart##1##2{}%
\let\hyper@linkend\@empty\citep[#1][#2]{#3}}}
\newcommandtwoopt{\citetads}[3][][]{\href{http://adsabs.harvard.edu/abs/#3}%
{\def\hyper@linkstart##1##2{}%
\let\hyper@linkend\@empty\citet[#1][#2]{#3}}}
\newcommandtwoopt{\citeyearads}[3][][]%
{\href{http://adsabs.harvard.edu/abs/#3}
{\def\hyper@linkstart##1##2{}%
\let\hyper@linkend\@empty\citeyear[#1][#2]{#3}}}
\makeatother



\newcommand{\adeg}[1]{{#1}$^{\circ}$}

\newcommand{\asec}[1]{{#1}$^{\prime\prime}$}

\newcommand{\mjy}[1]{{#1} mJy}

\newcommand{\ujybeam}[1]{{#1} $\mu$Jy beam$^{-1}$}

\newcommand{\bootes}{Bo\"otes~}

\definecolor{mygreen}{rgb}{0.0, 0.42, 0.0}

\usepackage[mathlines,switch]{lineno}

\newcommand{\ncommon}{10196~}

\usepackage{lipsum}

\newcommand\blfootnote[1]{%
  \begingroup
  \renewcommand\thefootnote{}\footnote{#1}%
  \addtocounter{footnote}{-1}%
  \endgroup
}

\graphicspath{{./fig/}}

\begin{document} 

    \title{Spectral indices in active galactic nuclei \\ as seen by Apertif and LOFAR}
    
    \titlerunning{Apertif deep fields release}    
    \authorrunning{A.\,Kutkin et al.}
   \author{
            A.\,M.\,Kutkin
            \inst{1^*}
        \and
            R.\,Morganti
            \inst{1,2}
        \and
            T.\,A.\,Oosterloo
            \inst{1,2}
        \and
            \\
            E.\,A.\,K.\,Adams
            \inst{1,2}
        \and
            H.\,D\'{e}nes
            \inst{3}
        \and
            J.\,van\,Leeuwen
            \inst{1}
        \and
            M.\,J.\,Norden
            \inst{1}
        \and
            E.\,Orru
            \inst{1}
        }
          
    \institute{
        ASTRON, The Netherlands Institute for Radio Astronomy, Oude Hoogeveensedijk 4, 7991 PD, Dwingeloo, The Netherlands
    \and
        Kapteyn Astronomical Institute, P.O. Box 800, 9700 AV Groningen, The Netherlands
    \and
        School of Physical Sciences and Nanotechnology, Yachay Tech University, Hacienda San Jos\'{e} S/N, 100119, Urcuqu\'{i}, Ecuador
    }
 
  \abstract
  {
We present two new radio continuum images obtained with Apertif at 1.4\,GHz. The images, produced with a direction-dependent calibration pipeline, cover 136 square degrees of the Lockman Hole and 24 square degrees of the ELAIS-N fields, with an average resolution of 17$\times$12\asec{} and residual noise of \ujybeam{33}. With the improved depth of the images we found  in total 63692 radio sources, many of which are detected for the first time at this frequency. With the addition of the previously published Apertif catalog for the \bootes field, we cross-match with the LOFAR deep-fields value-added catalogs at 150\,MHz, resulting in a homogeneous sample of \ncommon common sources with spectral index estimates, one of the largest to date.
We analyze and discuss the correlations between spectral index, redshift, linear sources size, and radio luminosity, taking into account biases of flux-density-limited surveys.  
Our results suggest that the observed correlation between spectral index and redshift of active galactic nuclei can be attributed to the Malmquist bias reflecting an intrinsic relation between radio luminosity and the spectral index. 
We also find a correlation between spectral index and linear source size with more compact sources having steeper spectra. 
}


\maketitle \blfootnote{\noindent Tables~\ref{tab:catalog} and \ref{tab:si} are only available in electronic form at \url{http://vo.astron.nl} and at the CDS via anonymous ftp to cdsarc.u-strasbg.fr (130.79.128.5) or via \url{http://cdsarc.u-strasbg.fr}}
\blfootnote{\noindent *kutkin@astron.nl}


\begin{figure*}[!h] 
    \centering
    \includegraphics[width=0.95\linewidth]{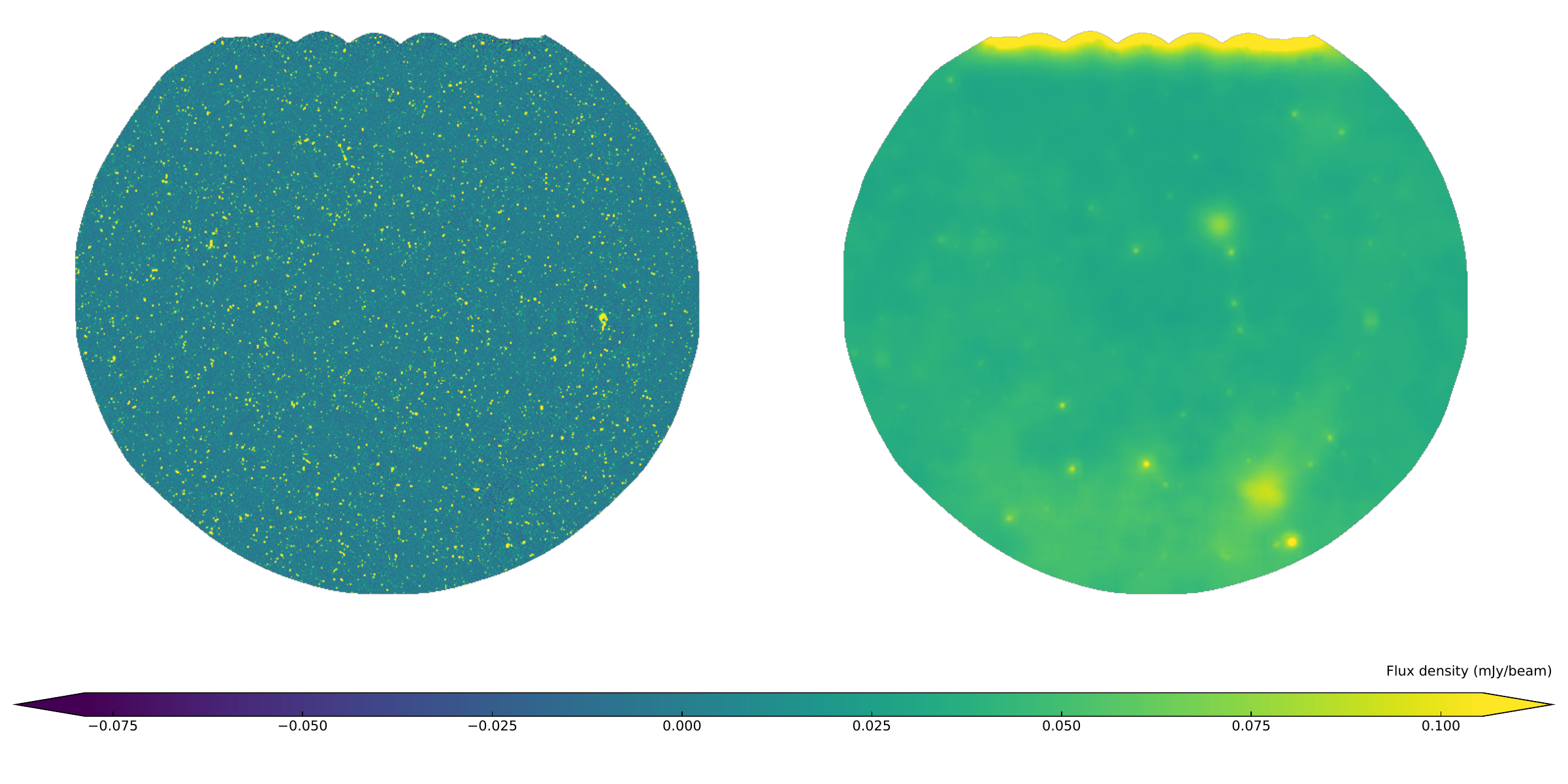}
    \includegraphics[width=0.95\linewidth]{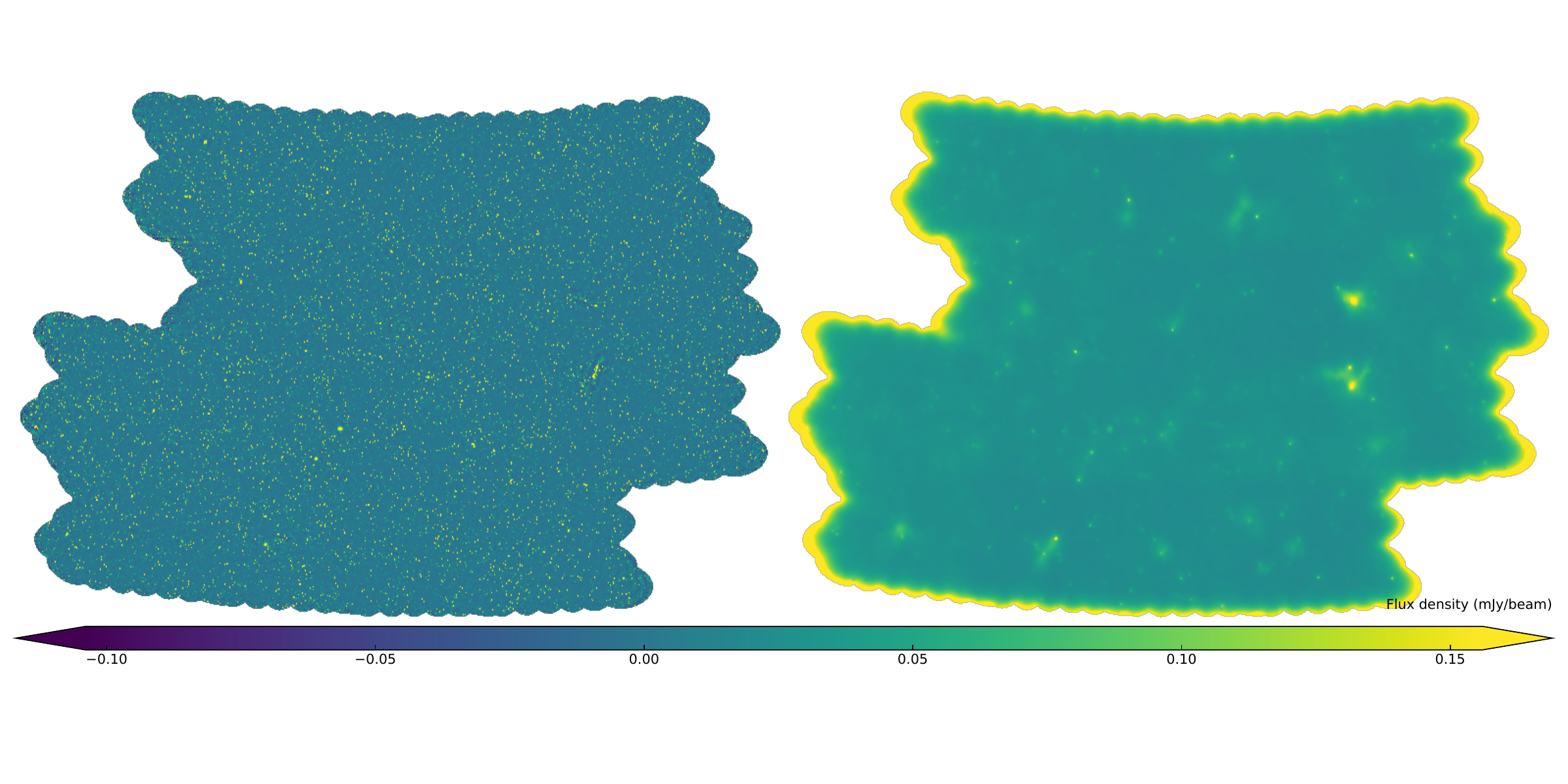}
    \caption{Mosaic images of the ELAIS-N (\textit{top}) and Lockman Hole (\textit{bottom}) fields along with the noise RMS maps.}
    \label{fig:mosaics}
\end{figure*}

\section{Introduction}
\label{sec:intro}

The shape of the radio spectral energy distribution of active galactic nuclei (AGN) carries information about the emission mechanisms, the physical properties of the sources, and their evolutionary stage \citep[][and references therein]{2017NatAs...1..596M, 2020NewAR..8801539H}. The radio spectrum from synchrotron emission is often well-described by a power law, characterized by the spectral index $\alpha$, defined as $S_\nu \propto \nu^\alpha$, where $S_\nu$ is the flux density at frequency $\nu$. Deviations from this power law are clear signatures of e.g. energy losses or absorption process that can be observed at high and low frequencies respectively. These features move in frequency not only with the age but also with the redshift of the source, as nicely illustrated by the sketch in Fig. 2 of \citet{2012MNRAS.420.2644K}.

Spectral indices of radio sources typically range from $-1.5$ to $0.5$, with the majority falling between $-0.8$ and $-0.7$ \citep{1992ARA&A..30..575C, 2016MNRAS.463.2997M}. Steep spectrum sources ($\alpha < -0.7$) are often associated with extended radio lobes, while flat spectrum sources ($-0.5 < \alpha < 0$) are generally compact and may indicate self-absorbed synchrotron emission~\citep[e.g.,][]{1966ApJ...146..621K}. Inverted spectrum sources ($\alpha > 0$) are less common and may be associated with young or highly compact absorbed objects~\citep{1998PASP..110..493O,2021A&ARv..29....3O}.

A relation between spectral index and redshift has been noted already in the early studies of radio sources~\citep{1979A&A....80...13B}, with higher redshift sources tending to have steeper spectra. This relation  played a key role in discovering the first high redshift galaxies, which, because of this, were all radio galaxies~\citep[e.g.][]{1994A&AS..108...79R, 2008A&ARv..15...67M}. 

As summarized by \cite{2008A&ARv..15...67M}, the relation that more distant sources have steeper spectra was originally interpreted as due to a combination of a concave radio spectrum (resulting from synchrotron and inverse Compton (IC) energy losses at high frequencies) combined with the k-correction, in which the steeper part of the radio spectrum enters the typical observed range of frequencies for high-$z$ sources. However, \cite{2006MNRAS.371..852K} shower a lack of strong curvature in the spectra of their sources, in direct contradiction of this explanation.

Alternative explanations for this relation have also been proposed. They include enhanced spectral aging due to increased IC losses against the cosmic microwave background (CMB) radiation at high redshifts \citep[][and references therein]{2006MNRAS.371..852K, 2018MNRAS.480.2726M};
intrinsic correlation between low-frequency spectral index and radio luminosity, with larger luminosities corresponding to larger jet powers which produce larger magnetic fields, with the end result of more rapid electron cooling times~\citep{1990ApJ...363...21C, 1999AJ....117..677B}; and
higher ambient density at high redshift which can  cause steeper electron energy spectra~\citep{1998JApA...19...63A,2006MNRAS.371..852K}.

The study of the spectral properties for large samples of radio sources is now facilitated by large-scale radio surveys across different frequencies~\citep[see][for the surveys overview]{Norris17}.
The LOFAR Two-meter Sky Survey~\citep[LoTSS;][]{2017A&A...598A.104S, 2022A&A...659A...1S}, with unprecedented sensitivity and resolution, has extended the detailed study of the spectral index to low frequencies. Furthermore, the availability of value-added catalogs for the LOFAR sources, which crucially include redshifts, make them extra suited for tracing the evolution of their properties.
Now, a bottle neck for probing spectral indices is availability of a survey at GHz frequencies deep enough to match the sensitivity of LOFAR. Shallow 1.4\,GHz surveys can bias samples towards flatter spectrum sources at higher redshifts \citep{2012MNRAS.420.2644K} while the deep surveys are usually focused on small sky areas lacking sources for statistical studies. 
This has placed the onus on higher frequency surveys to match the sensitivity and sky coverage of LOFAR. One example is the 1.4\,GHZ continuum survey with Apertif, presented and released partially in this paper.

Recent study by~\citet[][K23]{K23} probed spectral indices within the 26.5 square degree region of \bootes constellation using of LOFAR HBA 150\,MHz and LBA at 50~MHz as well as Apertif images at 1.4\,GHz. They found that the redshift dependence is different for the low- (50--150\,MHz) and high-frequency (150--1400\,MHz) spectral index. The first drops steeply at lower redshifts and decreases more slowly at higher redshifts, while the second only shows a weak linear trend throughout the entire redshift range. This can be explained by a population of peaked spectrum sources appearing at lower frequencies. In another recent work, \cite{2025MNRAS.tmp..278P}, authors claim that the correlation between spectral index and redshift is driven by the dependence of the former on luminosity and a strong redshift-luminosity correlation occurring in flux density-limited samples, aka the Malmquist bias. 
This naturally motivates further work with expanded sample sizes, especially at high redshifts, and more detailed investigation of any correlation of spectral index with other parameters, e.g. luminosity and linear source size. 

In this work we take this next step by examining spectral indices between 1.4\,GHz and 150\,MHz over almost 200 square degrees.
The paper is organized as follows. In Section~\ref{sec:data_apertif} the new 1.4 GHz images and catalogs obtained with Apertif are presented. In Section~\ref{sec:lofar} the combination of LOFAR and Apertif data is described. The analysis of spectral index and other source parameters is presented in Section~\ref{sec:analysis}. The results and conclusions are discussed in Section~\ref{sec:conclusions}. 

Through this paper we assume a flat $\Lambda$CDM cosmology model with $H_0=70$\,km/(s\,Mpc), $\Omega_M=0.3$, implemented by the \texttt{Astropy Python} library~\citep{astropy:2018}. We use a positively defined spectral index $\alpha=d\ln S/d\ln\nu$.

\section{Apertif images and catalogs}
\label{sec:data_apertif}

\begin{figure} 
    \centering
    \includegraphics[width=0.98\linewidth]{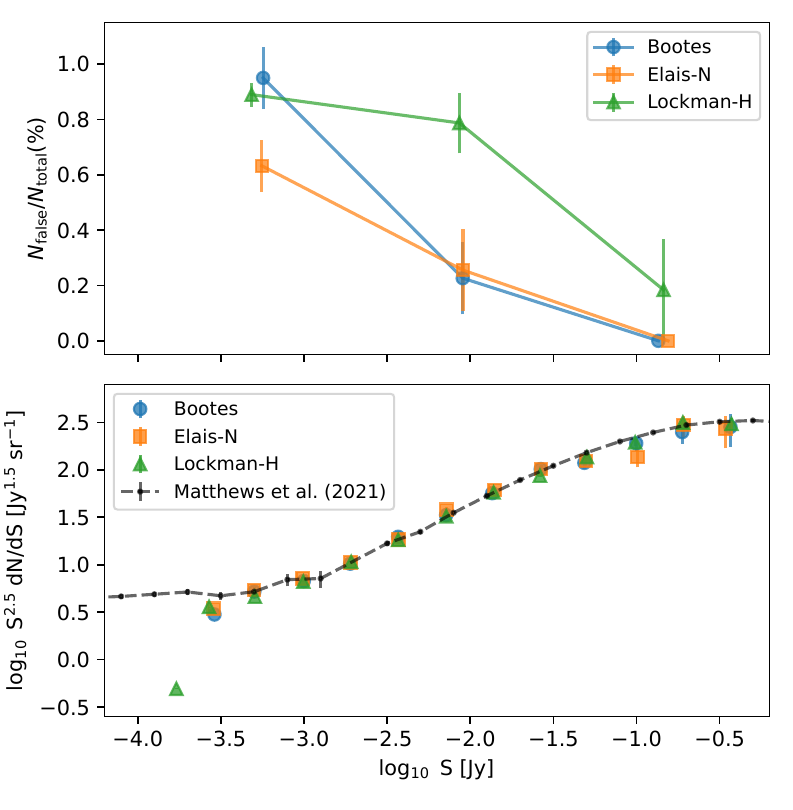}
    \caption{False-detection fraction and differential source counts in the Apertif catalog.}
    \label{fig:completeness}
\end{figure}

Apertif is a phased-array feed system that was mounted on the Westerbork Synthesis Radio Telescope (WSRT), providing a field-of-view of 10 sq deg \citep{2022A&A...658A.146V}.
As described in \cite{2022A&A...658A.146V} Apertif is a phased array feed system that was mounted on the Westerbork Synthesis Radio telescope and 
Apertif enabled large-scale radio continuum, polarization and spectral line surveys, which were supported by automated processing with a custom pipeline \citep{2022A&A...667A..38A}.

In \citet[][K22]{K22} the radio continuum images and catalog from the first data release are described. More recently, the pipeline for producing radio continuum images has been improved with the addition of direction-dependent calibration, as  described in K23, where the improved images for the Bo\"otes region were presented. 
Here we expand on this, by presenting the mosaic images of two more famous regions of the northern sky: ELAIS-N and Lockman Hole, produced with the new pipeline. 

The reduction and imaging of the data was done following the same procedure used for the Boötes field. The individual compound beam images produced with the improved pipeline were corrected for an astrometric offset and primary beam shape, brought to the same resolution, re-projected to a common sky grid and linearly mosaicked into a larger image (see K22, K23 for details). 
Because of the individual Apertif images overlap, the mosaicking leads to an increase in sensitivity. 
The mosaic images of the ELAIS-N and Lockman Hole fields along with the corresponding residual noise maps are shown in Figure~\ref{fig:mosaics}. The images and the catalogs are published at ASTRON VO server at \url{https://vo.astron.nl}. Example of the catalog records for the ELAIS-N field is shown in the Table~\ref{tab:catalog}.  

The corresponding source catalogs were extracted for these two fields using \texttt{PyBDSF} source finder~\citep{2015ascl.soft02007M}, in the same way as done for the Bo\"otes field by K23. The flux density and coordinate uncertainties were corrected for the systematic calibration errors. 
The false detection rate, below 2\%, was estimated as the number of sources detected in the inverted image over the number of sources in the original image. The completeness of all three catalogs was estimated to be \mjy{0.3} by comparing the differential source counts to \cite{2021ApJ...909..193M}. This is illustrated in Figure~\ref{fig:completeness} (see also K22 and K23 for details). Note, that the completeness and reliability estimates may have local variations following the changes of local noise RMS (see Figure~\ref{fig:mosaics}). 

For the further analysis we combine the catalogs of all three fields together, summarized in Table~\ref{tab:adata}.
We note, that individual Apertif images were obtained with a slightly different central frequency and bandwidth (1350\,MHz/150\,MHz before January 2021 and 1375\,MHz/250\,MHz after). For the new frequency setup, we have derived the corresponding primary beam models, which were taken into account during the mosaicing procedure. In this work we did not trace the primary beam shape variations with any higher cadence, which might be a reason for some systematic errors. However, since the mosaics are produced using more than one compound beam image taken at different epochs, we expect these errors to average out and to be well below the general calibration ones. For the further analysis, 
we calculate a median offset between Apertif and NVSS total flux density and correct the former by this factor, which is within a few percent for all three catalogs. We accordingly adopt the central frequency 1400\,MHz.

\begin{table*}
    \centering{
    \caption{Apertif source list. \label{tab:catalog}}
   {\setlength{\tabcolsep}{5pt}
    \begin{tabular}{ccccccccccc}
    \hline\hline
    Name & RA & $\sigma_\mathrm{RA}$ & Dec & $\sigma_\mathrm{Dec}$ & $S_\mathrm{total}$ &  $\sigma_{S_\mathrm{total}}$ &  $S_\mathrm{peak}$ &  $\sigma_{S_\mathrm{peak}}$ &  RMS & S\_Code \\
    ~  & [\adeg{}] & [\asec{}] & [\adeg{}] & [\asec{}] & mJy & mJy & mJy/bm & mJy/bm & $\muup$Jy/bm &  (S/M/C) \\
    (1) & (2) & (3) & (4) & (5) & (6) & (7) & (8) & (9) & (10) & (11) \\
     \hline
APTF\_J155225+545513 & 238.1062 & 1.27 & 54.9205 & 1.27 & 0.745 & 0.078 & 0.623 & 0.046 & 34.5 & S    \\
APTF\_J155225+550716 & 238.1076 & 1.29 & 55.1213 & 1.33 & 0.601 & 0.068 & 0.543 & 0.041 & 32.6 & S    \\
APTF\_J155228+541900 & 238.1175 & 1.40 & 54.3169 & 1.46 & 0.488 & 0.067 & 0.471 & 0.041 & 35.1 & S    \\
APTF\_J155229+545859 & 238.1211 & 1.95 & 54.9833 & 2.28 & 0.278 & 0.068 & 0.250 & 0.037 & 35.1 & S    \\
APTF\_J155229+542828 & 238.1228 & 1.85 & 54.4746 & 2.39 & 0.273 & 0.065 & 0.256 & 0.036 & 35.3 & S    \\
APTF\_J155230+555932 & 238.1252 & 2.20 & 55.9925 & 2.68 & 0.216 & 0.063 & 0.211 & 0.036 & 35.5 & S    \\
    \hline
    \end{tabular}
    }
    \tablefoot{Sample of the source list records for the ELAIS-N field. The columns are described in Sect.~\ref{sec:data_apertif}. The full tables for both fields containing are available in machine-readable format at \url{https://vo.astron.nl} and through the CDS.}
    }
\end{table*}


\begin{table*}[!h]
   \centering{
    \caption{Apertif data}
    \begin{tabular}{lcccccc}
    \hline\hline
        Field name & RA & DEC & Area & Resolution & Image noise & N sources\\
                   &    &     & [sq. deg.] & \asec{} & [\ujybeam{}] &  \\
        ~~~~~~(1) & (2) & (3) & (4) & (5) & (6) & (7) \\
    \hline     
         \bootes      & 14:31:53 & +34:27:44 & 27 & 23$\times$11 & 40 & 8994 \\
         Elais North  & 16:10:20 & +54:38:13 & 24 & 18$\times$11 & 37 & 8526 \\
         Lockman Hole & 10:57:40 & +57:02:36 & 136 & 15$\times$13 & 27 & 55166 \\
    \hline
    \end{tabular}
    \tablefoot{Columns designation: (1): Field name (2--3): Approximate central coordinates; (4): Image area (square degrees) (5): Image resolution (6): Median value of the background noise RMS reported by PyBDSF (\texttt{Isl\_rms}) (7): Number of sources in the corresponding catalog}
    \label{tab:adata}
   }
\end{table*}


\section{Combination with LOFAR}
\label{sec:lofar}

The same deep fields were observed by LOFAR and an extensive data analysis has been performed on these observations~\citep{2021A&A...648A...1T, 2021A&A...648A...2S}. 
As demonstrated by K22 and K23, combination between the Apertif and LOFAR surveys (at 1400\,MHz and 150\,MHz) provides a great synergy due to their common sky coverage, similar angular resolution and high sensitivity. This combination allows probing spectral indexes of the sources at flux densities and redshifts not reachable before. 

We used the LOFAR value-added deep fileds catalogs available through the official LOFAR web-site~\footnote{https://www.lofar-surveys.org/}~\citep{2019A&A...622A...3D, 2021A&A...648A...3K}. As a redshift estimate we used column labeled `Z\_BEST' which refers to the best available redshift estimate, either spectroscopic or photometric. 

We cross-matched the source coordinates between Apertif and LOFAR samples using a \asec{5} radius and obtained a sample of \ncommon sources, 96\% of which have redshift measurements. 
The spectral indices were estimated as: $\alpha^{1400}_{150}=\log\,(S^{1400}_{\rm int}/S^{150}_{\rm int})/\log\,(1400/150)$, where $S^{1400}_{\rm int}$ and $S^{150}_{\rm int}$ are the Apertif and LOFAR integrated flux density. The spectral index error estimate is propagated from flux density uncertainties. These data are summarized in Table~\ref{tab:si} for all \ncommon common sources.

\begin{table*}[]
   \centering{
    \caption{Common Apertif/LOFAR sources}
    \begin{tabular}{llccccc}
    \hline\hline\smallskip
    Apertif name & LOFAR name & $\alpha_{150}^{1400}$ & $\sigma_{\alpha}$ & $z$ & $\log\,L_\nu^{150}$ & $D^{150}$ \\
    &    &     &  &  & [W\,Hz$^{-1}]$ & [kpc] \\
        ~~~~~~(1) & (2) & (3) & (4) & (5) & (6) & (7)\\
    \hline         
 APTF\_J143808+334640 & ILTJ143808.71+334641.2 & -0.60 & 0.13 & 1.30 & 24.3 & 23.53 \\
APTF\_J143812+343414 & ILTJ143811.96+343416.4 & -0.15 & 0.16 & 1.55 & 24.2 &  0.00 \\
APTF\_J143812+344012 & ILTJ143812.31+344012.3 & -0.65 & 0.07 & 0.13 & 22.8 &  2.60 \\
APTF\_J143814+352808 & ILTJ143814.81+352808.2 & -0.86 & 0.02 & 1.53 & 26.2 & 94.98 \\
APTF\_J143811+343104 & ILTJ143811.18+343104.6 & -0.43 & 0.11 & 0.68 & 23.8 &  0.00 \\
APTF\_J143811+343338 & ILTJ143811.26+343341.3 & -0.92 & 0.12 & 3.83 & 25.0 &  0.00 \\
    \hline
    \end{tabular}
\tablefoot{Columns designation: (1--2): Apertif and LOFAR source name (J2000), (3-4): Spectral index and its error, (5): Redshift (6): Log of luminosity density at 150\,MHz (7): Linear size at 150\,MHz. The full table containing \ncommon entries is available in machine-readable format at \url{https://vo.astron.nl} and through the CDS.}  \label{tab:si}
}
\end{table*}


\section{Analysis and results}
\label{sec:analysis}

\subsection{Spectral index and total flux density}
\label{sec:si_flux}

\begin{figure}[ht!]
    \centering
    \includegraphics[width=1.01\linewidth]{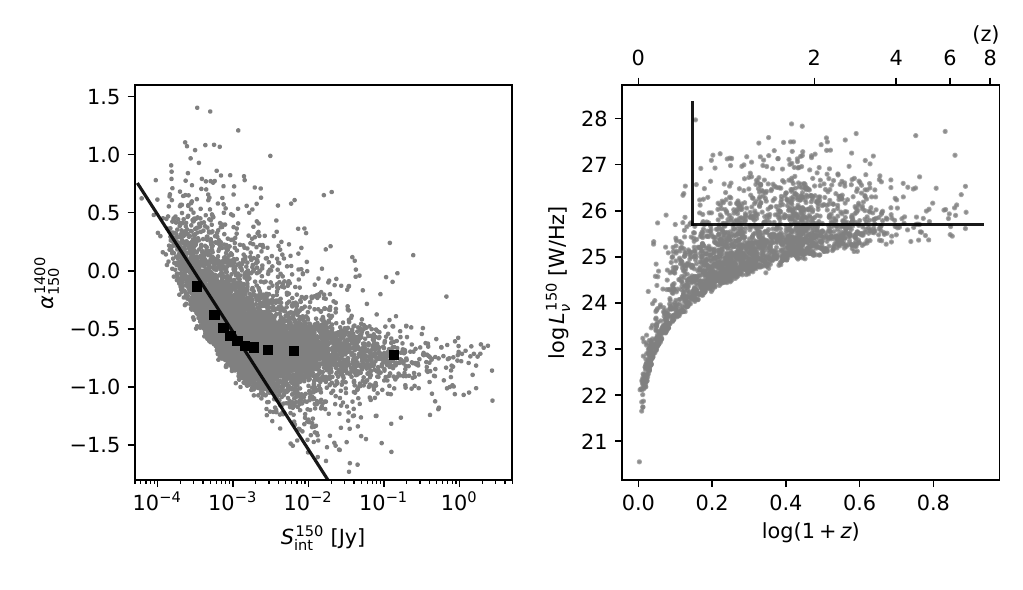}
    \caption{\textit{Left:} Spectral index against the LOFAR integrated flux density for all common Apertif/LOFAR sources. Markers show median spectral index in a given flux density interval. Solid line indicates a lower limit calculated for the Apertif completeness level of 0.3\,mJy (Sec~\ref{sec:si_flux}). \textit{Right}:Luminosity against redshift for the main sample. Solid lines indicate a Malmquist bias free region (Sec~\ref{sec:SI_lum}).
    }
    \label{fig:bias}
\end{figure}

As mentioned above, the Apertif catalog is complete down to 0.3\,mJy level. Below this level a source may be missing in the Apertif catalog while detected by LOFAR, unless it has a flat spectrum. 
This results in a bias of the average spectral index estimate for sources with low flux density, as shown in the left panel of Figure~\ref{fig:bias} where the distribution of spectral indices as a function of LOFAR integrated flux density is plotted for the common sources. It is clear that below a few mJy the spectral index distribution is affected by incompleteness due missing steep spectrum sources in the Apertif survey. To exclude this bias, we limit the further analysis to the threshold $S_{int}^{150} > 3$\,mJy, where the spectral index distribution becomes symmetric and its median value does not change significantly with the flux density (see left plot of Fig.~\ref{fig:bias}). As a result we obtain a sample of 2437 sources out of which 2273 have redshift estimates, further referred as main sample. Such a selection also implies that the great majority of the considered sources are radio loud AGN, which dominate the population at LOFAR flux densities above \mjy{3}~\citep[see Fig.\,10 and discussion by][]{2023MNRAS.523.1729B}. 


\subsection{Spectral index and luminosity}
\label{sec:SI_lum}

\begin{figure*}[!ht]
    \centering
    \includegraphics[width=0.99\linewidth] {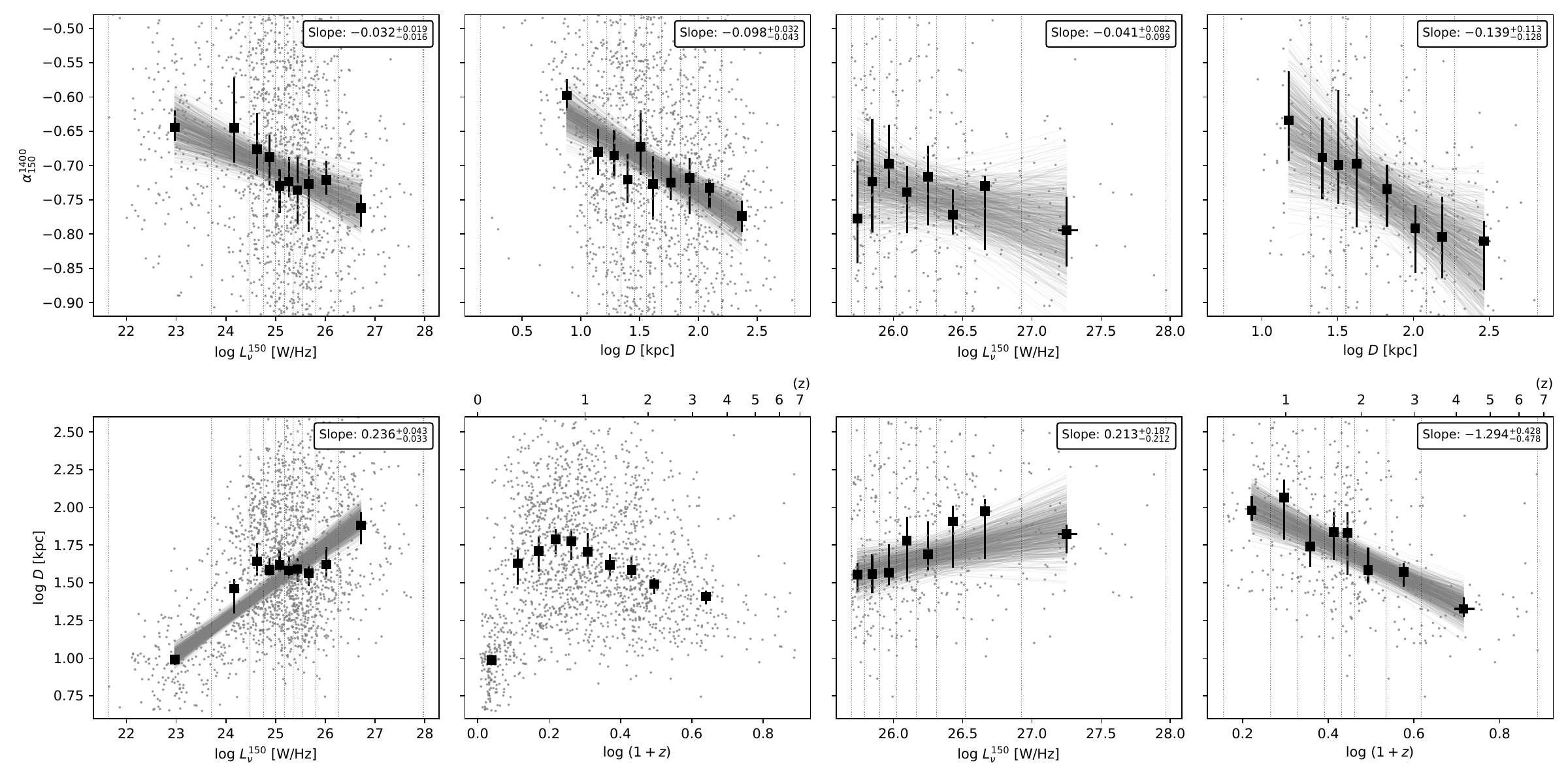}
    \caption{\textit{Top row}: Spectral index vs luminosity and linear source size for the main sample (\textit{columns 1--2}) and the Malmquist-bias free subsample (\textit{columns 3--4}). \textit{Bottom row}: Linear source size against luminosity and redshift for the same samples (see Sec~\ref{sec:SI_lum} and \ref{sec:size} for details).
    }
    \label{fig:correlations}
\end{figure*}

We calculate the radio luminosity density in the source frame as:
$
    L_\nu = 4\pi D_{\rm L}^2 S^\nu_{int}(1+z)^{-(1+\alpha)},
$
where $D_{\rm L}$ is the luminosity distance and $S^\nu_{int}$ the observed total flux density~\citep[e.g.,][]{2006MNRAS.371..852K}. 

In the right panel of Figure~\ref{fig:bias} the radio luminosity is plotted against  redshift for the main sample. Unsurprisingly,  detectable source luminosities depend strongly on the redshift,  resulting in a Malmquist bias when considering sources properties at different redshifts. In order to account for this bias, we subsequently consider a subsample of sources with luminosities above $5\cdot10^{25}$\,W/Hz and redshifts above 0.4, shown with solid lines in the right panel of Figure~\ref{fig:bias}. This includes 543 sources and shows no correlation between luminosity and redshift. 

In the top left panel of Figure~\ref{fig:correlations}, the distribution of spectral index against luminosity is shown for the main sample. In order to trace the correlation, we bin the X-axis into bins containing equal amount of sources;  the edges of these bins are shown with vertical dashed lines. Within each bin we calculate the median spectral index and average X-coordinate. These measurements are plotted with square markers along with their 95\% confidence intervals estimated using bootstrap. We then use MCMC simulations~\citep{2013PASP..125..306F} with a linear model $y=ax+b$ to determine the correlation coefficient, $a$, which is shown in top right corner. The solid gray lines show random samples drawn from an MCMC posterior distribution.

There is a weak correlation with more luminous sources showing steeper spectra following $\alpha \propto L^{-0.03}$. This correlation holds, with larger uncertainties, for the Malmquist-bias free subsample, as shown on the third panel of Figure~\ref{fig:correlations}.


\subsection{Spectral index and source size}
\label{sec:size}

Linear source size can be estimated as $D^{150}=\theta^{150} D_{\rm A}$, where $\theta^{150}$ is the LOFAR deconvolved source size (\texttt{PyBDSF: DC\_Maj}) and $D_{\rm A}$ is the angular diameter distance. To trace the correlation between the linear size and other parameters we consider only sources with $\theta^{150}>0$. 
As shown in the second panel of Figure~\ref{fig:correlations} there is a correlation with larger sources having steeper spectra following $\alpha \propto D^{-0.1}$. This trend holds if we consider the Malmquist bias free subsample (see Section~\ref{sec:SI_lum} and the fourth panel of Figure~\ref{fig:correlations}). 

In the second row of Figure~\ref{fig:correlations} the linear source size is plotted against luminosity and redshift for the main sample and the Malmquist bias free subsample. While there is an expected trend of more luminous sources to have larger sizes, the dependence of the size on the redshift is not monotonous~\citep[see also][for discussion]{2012MNRAS.420.2644K}.


\subsection{Spectral index and redshift}
\label{sec:SI_z}

\begin{figure}
    \centering
    \includegraphics[width=\linewidth]{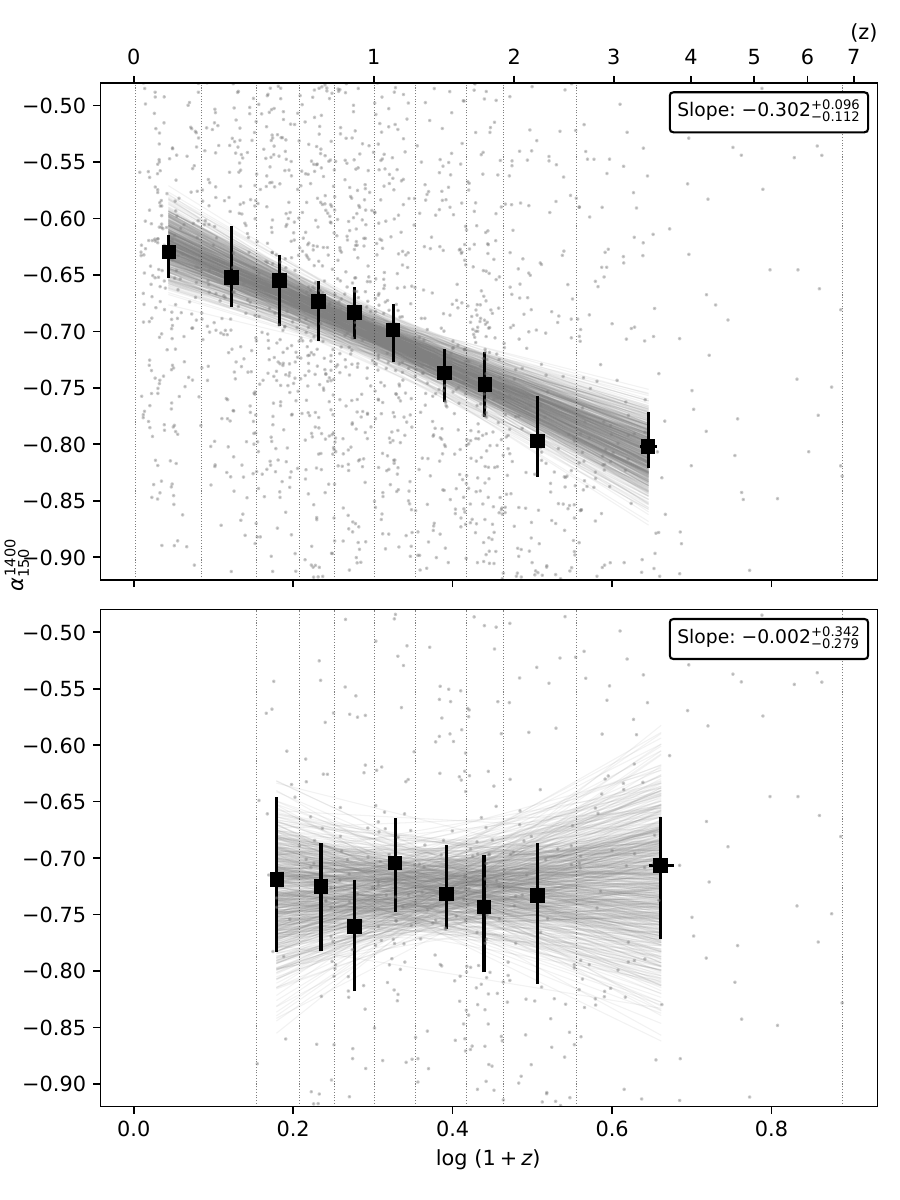}
    \caption{Spectral index against redshift for all sources (\textit{top}) and the Malmquist bias free subsample (\textit{bottom}).}
    \label{fig:si_z}
\end{figure}

\begin{figure}
    \centering
    \includegraphics[width=\linewidth]{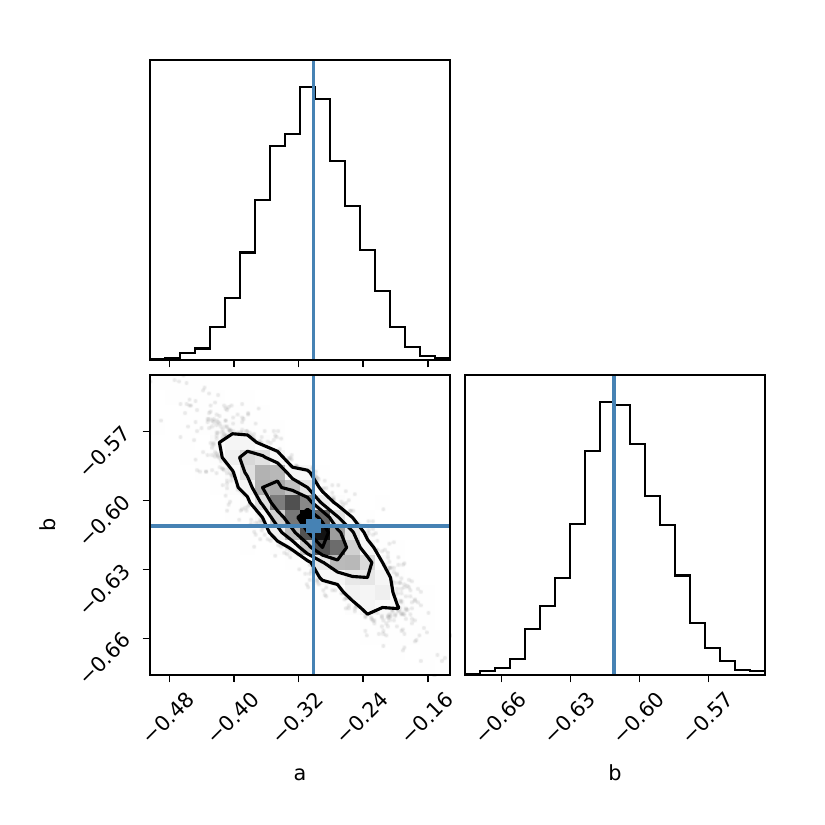}
    \caption{Parameters distributions of the MCMC fit with a linear model $\alpha(z)=a\,z+b$ shown at the top panel of Figure~\ref{fig:si_z}. Vertical and horizontal lines show the maximum a posteriori (MAP) estimate of the parameters.}
    \label{fig:ccplot}
\end{figure}

The distribution of spectral indices against redshift is presented in Figure~\ref{fig:si_z}. The individual sources are shown with gray dots. 
Binning is done in the same way as described before for the Figure~\ref{fig:correlations}. For each interval we calculate average redshift and median spectral index plotted with square markers. The uncertainties are estimated using bootstrap performed in each bin. 
The two plots present the trend for the main sample and for the Malmquist bias free subsample, plotted with the same binning. The slope coefficients and their errors obtained using a linear model and MCMC simulations are illustrated in Figure~\ref{fig:ccplot}. 

The top plot shows a clear correlation with steeper spectra at high redshift. The slope, $\alpha \propto \log(1+z)^{-0.3\pm0.1}$ obtained for the main sample agrees within errors with the results of K23. As mentioned in the introduction this well-known trend has been used to select high-redshift sources based on their steep spectra. However, there is no $\alpha$-$z$ correlation for the Malmquist-bias free subsample as seen in the bottom panel of Figure~\ref{fig:si_z} (see also Section~\ref{sec:SI_lum}). This implies that we see steeper spectra at higher redshifts mostly due to the combination of $\alpha$-$L$ correlation and the Malmquist bias.

\section{Discussion and Conclusions}
\label{sec:conclusions}

We have processed Apertif observations and presented two new radio continuum images and catalogs at 1.4\,GHz. Due to higher sensitivity of Apertif compared to previous northern surveys, many radio sources were detected for the first time at these frequencies. The images cover more than 150 square degrees within famous fields, ELAIS-N and Lockman Hole, observed with many other facilities at various frequencies. With the addition of previously published catalog for the \bootes filed, we cross-matched Apertif source coordinates with those of LOFAR value-added catalogs to probe spectral indices and other parameters of radio sources. 
We compile and study a new sample of common Apertif/LOFAR sources, complete in terms of spectral index and including enough data for the statistical analysis which is especially crucial for high redshift sources. We constrain and study separately a subsample of sources free from Malmquist bias.

The median spectral index of radio sources shows anti-correlation with luminosity and linear source size: the more powerful and more extended sources tend to have steeper spectra. The observed positive correlation between the size and the luminosity agrees well with this trend. 
The observed correlation between spectral index and redshift can be attributed to the Malmquist bias and the intrinsic $\alpha$-$L$ correlation. This can be explained with a model proposed by~\cite{1962SvA.....6..317K}, where an optically thin part of a radio spectrum experience a break from $\alpha_0$ to $\alpha_0 - 0.5$ when the radiation losses timescale exceeds the injection one. The break frequency is inversely proportional to the jet magnetic field strength, $\nu_{\rm break}\propto B^{-3}$, while the total radio luminosity is proportional to $B^{3.5}$. Therefore, more luminous sources will have the spectral break at lower frequencies~\citep[see e.g.][for details]{1990ApJ...363...21C}. 

After removing the Malmquist bias, there is still a significant anti-correlation between $\alpha$-$D$ and $D$-$z$. The first correlation can be explained by source aging, when older sources which had enough time to grow to larger sizes have steeper spectra. The second one might be related to younger, hence smaller, sources seen at higher redshifts. 
At the same time there is no correlation of $\alpha$-$z$ seen after eliminating the Malmquist bias, implying that there is another mechanism responsible for steepening of radio spectra at higher redshift not related to $\alpha(L,D)$. This can be a more efficient electron cooling through the inverse Compton losses due to more dense environment at higher redshifts. Thereby, we can speculate about two concurring trends eliminating the $\alpha$-$z$ correlation: IC-losses cause spectra steeper while a population of compact sources with flatter spectra arises at higher redshifts~\citep[see][for more discussion]{2012MNRAS.420.2644K,2018MNRAS.480.2726M}.

Interestingly, \cite{2012MNRAS.420.2644K} who used combined smaller samples reported a weak $\alpha$-$z$ correlation even after removal of the Malmquist bias. 
We confirm a slight insignificant trend with a negative slope if the sample is constrained to have $\alpha<-0.5$, as performed in the mentioned work. The authors try to exclude a population of bright sources with flatter spectra appearing at higher redshifts represented by compact BL\,Lacertae objects and flat spectrum radio quasars. In this work we consider a sample complete in terms of spectral index since the distribution of the latter is continuous at any redshift range. 

The intrinsic scatter in spectral index distribution and weak $\alpha$-$z$ correlation even within the total biased sample do not allow a reliable high-redshift sources identification by their steep spectra only. More robust results can be achieved involving other parameters, like a linear source size. 

Apertif with its angular resolution, sensitivity and sky coverage excellently complements the LOFAR observations allowing not only statistical studies of radio spectra, but also probe internal structures within extended sources. We continue processing the observations and release new images for the community.


\begin{acknowledgements}

This work makes use of data from the Apertif system installed at the Westerbork Synthesis Radio Telescope owned by ASTRON. ASTRON, the Netherlands Institute for Radio Astronomy, is an institute of the Dutch Science Organisation (De Nederlandse Organisatie voor Wetenschappelijk Onderzoek, NWO). Apertif was partly financed by the NWO Groot projects Apertif (175.010.2005.015) and Apropos (175.010.2009.012).
This research made use of \texttt{Python} programming language with its standard and external libraries/packages including \texttt{numpy}~\citep{numpy}, \texttt{scipy}~\citep{scipy}, \texttt{scikit-learn}~\citep{scikit-learn}, \texttt{matplotlib}~\citep{matplotlib}, \texttt{pandas}~\citep{pandas} etc.
This research made use of Astropy,\footnote{http://www.astropy.org} a community-developed core Python package for Astronomy \citep{astropy:2013, astropy:2018}. 
The \texttt{radio\_beam} and \texttt{reproject} python packages are used for manipulations with restoring beam and reprojecting/mosaicking of the images. 
This research has made use of "Aladin sky atlas" developed at CDS, Strasbourg Observatory, France~\citep{2000A&AS..143...33B, 2014ASPC..485..277B}. 

\end{acknowledgements}

\bibliographystyle{aa}
\bibliography{aa54375-25}

\end{document}